\documentclass[aps,twocolumn,showpacs,preprintnumbers,prb,amsmath,amssymb,amsfonts,superscriptaddress,floatfix]{revtex4-1}
\usepackage{graphicx}
\usepackage{dcolumn}
\usepackage{bm}
\usepackage[version=3]{mhchem} 
\usepackage{color}
\usepackage{array}
\usepackage{multirow}

\begin{document}
\title{Density Functional Methods for the Magnetism
of Transition Metals: SCAN in Relation to Other Functionals}

\author{Yuhao Fu}

\author{David J. Singh}
\email{singhdj@missouri.edu}

\affiliation{Department of Physics and Astronomy, University of Missouri, Columbia, MO 65211-7010 USA}
\date{\today}

\begin{abstract}
We report tests of various density functionals for ferromagnetic, Fe, Co and Ni
with a focus on characterizing the behavior of the
so-called strongly constrained and appropriately normed (SCAN) functional.
It is found that SCAN is closer in behavior to functionals that yield
localized behavior, such as hybrid functionals, than other semilocal
functionals that are tested. The results are understood in terms of
a tendency to differentiate orbitals, favoring integer occupation,
which is necessary for a correct description of atomic systems, but
inappropriate for the open shell metallic ferromagnetic metals studied here.
\end{abstract}

\maketitle

\section{Introduction}

The 3$d$ transition metals and their compounds present an exceptional range
of physical behavior in part because of the possibility for the 3$d$
electrons to be localized, itinerant or in between.
Examples of this rich physics include the band structure related magnetism
of elemental Fe, Co and Ni, the Mott insulating physics of many transition
metal oxides, and the apparently
distinct high temperature superconductivity of
cuprates and Fe-pnictides.
A long standing goal has been the development of predictive theories,
and in particular,
computationally tractable theories that can reliably capture the physics
of this range of materials.
The challenge to theory is to develop density functionals that can
describe both localized and itinerant behavior in a predictive way.
In this regard, the concept of ``Jacob's Ladder" has become
widely held. In this view, adding
more ingredients in density functionals, and constraining this
added flexibility by appropriate exact relations
should give overall more accurate descriptions of atoms, molecules
and solids.
\cite{perdew-05,kummel,tran-16}

It has long been recognized that functionals such as the local (spin)
density approximation (LDA) and standard generalized gradient approximations
(GGAs) do not provide an adequate treatment of correlated 3$d$ oxides
such as the Mott insulating parents of the cuprate superconductors.
\cite{pickett,singh-cacuo2}
This is due to an inadequate treatment of correlations, which can be
traced to self-interaction errors, \cite{svane}
and an insufficient tendency towards
integer orbital occupations. 
This is the basis for methods that add a Hubbard $U$ correction, i.e.
LDA+$U$. \cite{anisimov,anisimov1}
It is also a key aspect of a correct description of atoms and
molecules, where it can be traced to the need for a correct
description of the exchange correlation energy as a function of 
occupation number in order to avoid delocalization errors.
\cite{cohen,mori-sanchez,rivero}
These delocalization errors can also be reduced by hybrid functionals,
which can also correctly predict an antiferromagnetic insulating ground 
state for Mott insulators, including the high-T$_c$ parent, La$_2$CuO$_4$.
\cite{martin,feng,perry,rivero}
These incorporate explicit orbital dependence, and have been
very important in the
description of semiconductors, where in addition to improved
ground state properties they also improve band gaps.

Recently, a semi-local meta-GGA functional, including
the local (spin) densities, gradients and a kinetic energy density
was proposed. This is the so-called strongly constrained
and appropriately normed (SCAN) functional. \cite{sun,zhang}
Advantages of this meta-GGA approach include the fact that
calculations using a meta-GGA are computationally much less demanding
than hybrid functional calculations for solids,
the kinetic energy density can provide information about orbital
character, and calculations of energetics for diverse systems
showed that the SCAN functional provides improvements over
other functionals for a wide variety of solid state and molecular
properties.
\cite{zhang,sun2,tran,isaacs}
Significantly, based on calculations, unlike GGAs, this functional
is able to describe the ground state of La$_2$CuO$_4$ at least qualitatively.
\cite{lane}

Thus is reasonable to suppose that the SCAN functional 
provides an overall
improvement in the description of transition metals and compounds.
However, recent results, especially for Fe, Co and Ni,
show this not to be the case.
\cite{ekholm,jana,fu}
In particular, the magnetic tendency of these ferromagnetic metals
is strongly overestimated, as are the magnetic energies.
This is particularly severe in the case of Fe.
In that case the large errors make the SCAN functional incapable of
describing the phase stabilities underlying the materials science of steel.
\cite{fu}

Here we compare SCAN with other functionals for these transition metal
ferromagnets. The results show the origin of the problems in the SCAN
treatment of these materials is in the differentiation of orbitals.
This underscores the difficulty of finding a functional that can
treat the full range of transition metal magnets. Such a functional
must include the atomic physics favoring integer occupations in
Mott insulators, such as undoped cuprates,
and also the tendency towards multi-orbital open
shell behavior materials
that have significant itinerancy, such as Fe
and Fe-based superconductors. \cite{lu}
Thus the problem of developing a functional that properly describes
both localized electron behavior and itinerant behavior remains to be
solved.

\section{Computational Methods}

We performed first principles calculations
with several different exchange-correlation functionals.
These were the LDA,
the PBE GGA, \cite{pbe}
the SCAN, \cite{sun} TPSS \cite{tpss}
and revTPSS meta-GGAs, \cite{revtpss,revtpss2} and
the HSE03, \cite{hse03,hse03-2} HSE06 \cite{hse06}
and PBE0 \cite{pbe0,pbe0-2} hybrid functionals.
We additionally performed PBE+$U$ calculations with various choices of the
Hubbard parameter, $U$.

We used two different methods,
specifically the projector augmented wave (PAW) method
as implemented in the VASP code, \cite{PAW,VASP}
and the all electron general potential linearized augmented plane wave (LAPW)
method, \cite{singh-lapw} as implemented in the WINEN2K code. \cite{wien2k}
Here we focus
on results at the experimental lattice parameters in order to better
compare the description of magnetism in different approaches.
These are $a$=2.860 \AA, for bcc Fe, $a$=3.523 \AA, for fcc Ni
and $a$=2.507 \AA, $c$=4.070 \AA, for hcp Co.
The dependencies on lattice parameter for Fe comparing SCAN, PBE and LDA
were presented previously. \cite{fu}

The VASP and WIEN2k codes have different implementations of the meta-GGA
functionals, which involve different approximations. VASP implements
self-consistent calculations, but requires the use of PAW pseudopotentials,
which are not available for the meta-GGA functionals.
As such, we relied on PBE GGA PAW pseudopotentials, which is an approximation.
The general potential LAPW method, implemented in WIEN2k, is an
all-electron method, and does not use pseudopotentials.
Additionally, WIEN2k has an efficient implementation of the 
fixed spin moment constrained density functional theory method.
\cite{fsm1,fsm2}
This allows calculation of the magnetic energy as a function of the
imposed magnetization, which provides additional information about
the behavior of the functionals, and also facilitates determination
of the magnetic contribution to the energy.
However,
WIEN2k does not provide a self-consistent calculation with meta-GGA
potentials, and instead the energy is calculated for densities obtained
with the PBE GGA. This is an approximation for the energies, and
also prevents calculation of the electronic band structures with the
meta-GGA potentials.
We find that the approximations above are not significant for the
energies of the transition metals discussed here.
A comparison of the different methods for Fe is presented in
Table \ref{compare}.

The PAW calculations were done with a planewave
kinetic energy cut-off of 400 eV. We used converged, tested
Brillouin zone samplings, based on uniform meshes.
The LAPW calculations were done with sphere radii of $R$=2.25 Bohr and
basis sets defined by $RK_{max}$=9, where $K_{max}$ is the planewave
sector cut-off. Local orbitals were used to treat semicore states.
These are standard settings.

For the PBE+$U$ calculations, we used different values of $U$, with the
standard procedure of Dudarev and co-workers,
\cite{dudarev}
and the fully localized limit (SIC) double counting.
We covered the range up to $U$=8 eV.
It is common to find values in
range of $U$=5 eV to $U$=8 eV employed,
in literature for 3$d$ transition metal compounds.
This is based in part on the view that $U$ is a local atomic
quantity that describes correlation effects that are missing in
calculations with standard density functionals, such as the LDA or
PBE GGA, and therefore does not have much system dependence.

\begin{table}
\caption{Comparison of results for bcc Fe at the experimental
lattice parameters using
VASP and WIEN2k as employed here (see text).}
\begin{tabular}{llcccc}
\hline \hline
              &                         &  ~~LDA~~  & ~~PBE~~   & ~SCAN~ \\
\hline
VASP     & m$_{sp}$ ($\mu_B$)      & 2.15  & 2.25  & 2.65 \\
              & $\Delta E_{mag}$ (meV)~~~~ & -411 & -571 & -1081 \\
WIEN2k~~~~~~~ & m$_{sp}$ ($\mu_B$)     & 2.20  & 2.21  & 2.63 \\
              & $\Delta E_{mag}$ (meV) & -442 & -561 & -1114 \\
\hline
\end{tabular}
\label{compare}
\end{table}

\begin{figure}[h]
\includegraphics[width=0.95\columnwidth]{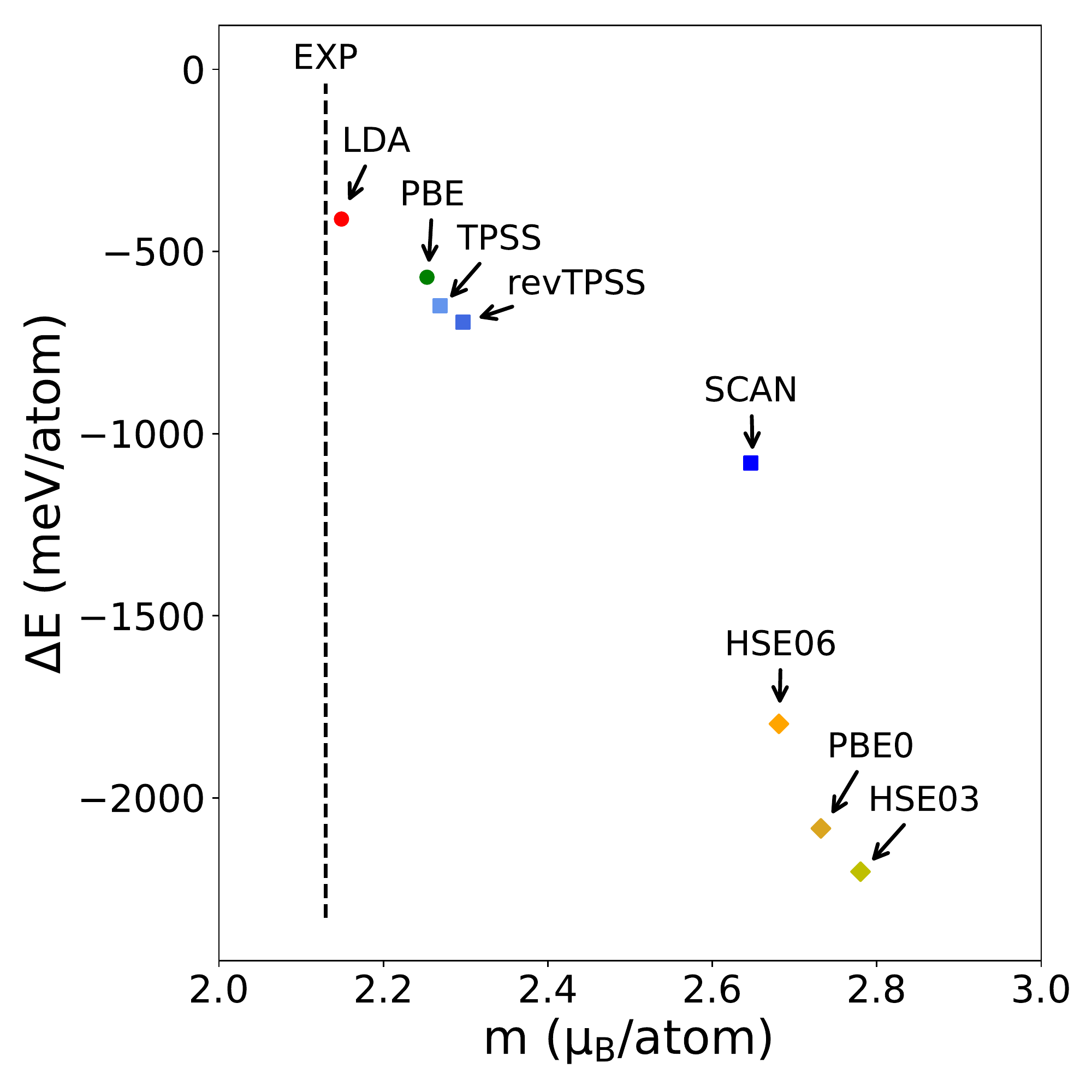}
\caption{The calculated magnetic energy and spin magnetizations of bcc Fe.
The magnetic energy is defined as the difference in energy between
the non-spin-polarized and ferromagnetic ground state.
The calculations were done self-consistently, using VASP for the
LDA, GGA
and meta-GGA functionals and WIEN2k for the hybrid functionals.}
\label{Fe-bcc-energy}
\end{figure}

This view has been criticized, however, and procedures have been
developed for determining $U$ for particular systems. These
are based on constrained calculations changing the electron occupation
of the orbital of interest, or by linear response.
\cite{himmetoglu,cococcioni}
This is important because screening, which strongly reduces $U$ from its
bare atomic value, varies from system to system.
For our purpose, it is important to note that these methods for obtaining
$U$ are designed to improve the description of the energy variation
between integer orbital occupations, i.e.
line segments connecting the values at integer occupations.

In any case, we also estimated values of $U$ using the linear response
method implemented in Quantum ESPRESSO. These estimated values
are 4.33 eV for Fe, 2.62 eV for Co and 5.56 eV for Ni.
While the basis set and implementation of PBE+$U$ in the LAPW method
and the pseudopotential planewave method of Quantum ESPRESSO are different,
so that values cannot be directly translated, these do provide an indication
of the rough values and we present also calculations for these specific
values.
Importantly, these values are never zero, although, as discussed below,
a value of zero gives the best agreement with experiment
for these elemental transition metal ferromagnets.

\section{Results and Discussion}

\begin{table*}
\caption{Properties of bcc Fe at its experimental 
lattice parameter ($a_{exp})$ and its calculated lattice
parameter ($a_{cal})$.}
\centering
\begin{tabular}{lccccccccc}
\hline\hline
                                                          &  LDA  & PBE   & TPSS & revTPSS & SCAN   & HSE06 & PBE0 & HSE03 & Expt. \\
\hline
$a_{cal}$ (\AA)                                               &  2.75  & 2.83 & 2.80  & 2.80       & 2.84     & ...         & ...        & ...         & 2.86 \\
m$_{sp}$ (a$_{exp}$) ($\mu_B$)      & 2.15  & 2.25  & 2.27  & 2.30       & 2.65     & 2.68     & 2.73   & 2.78     & 2.13\\
m$_{sp}$ (a$_{cal}$) ($\mu_B$)       & 1.93  & 2.20  & 2.15 & 2.16        & 2.57     & ...          & ...       & ...         & ... \\
$\Delta E_{mag}$ (a$_{exp}$) (meV) & -411 & -571 & -650  & -695       & -1081  & -1797  & -2202 & -2084 & ... \\
$\Delta E_{mag}$ (a$_{cal}$) (meV)  & -279 & -528 & -567  & -598       & -1028  & ...          & ...       & ...         & \\
\hline
\end{tabular}
\label{Fe-bcc-energy-v}
\end{table*}

We begin with bcc Fe.
Figure \ref{Fe-bcc-energy} shows our results for the magnetic energy of bcc Fe
with different exchange-correlation functionals,
at its experimental lattice parameter.
Numerical values and magnetic moments are given in Table \ref{Fe-bcc-energy-v}.

\begin{figure}[h]
\includegraphics[width=0.95\columnwidth]{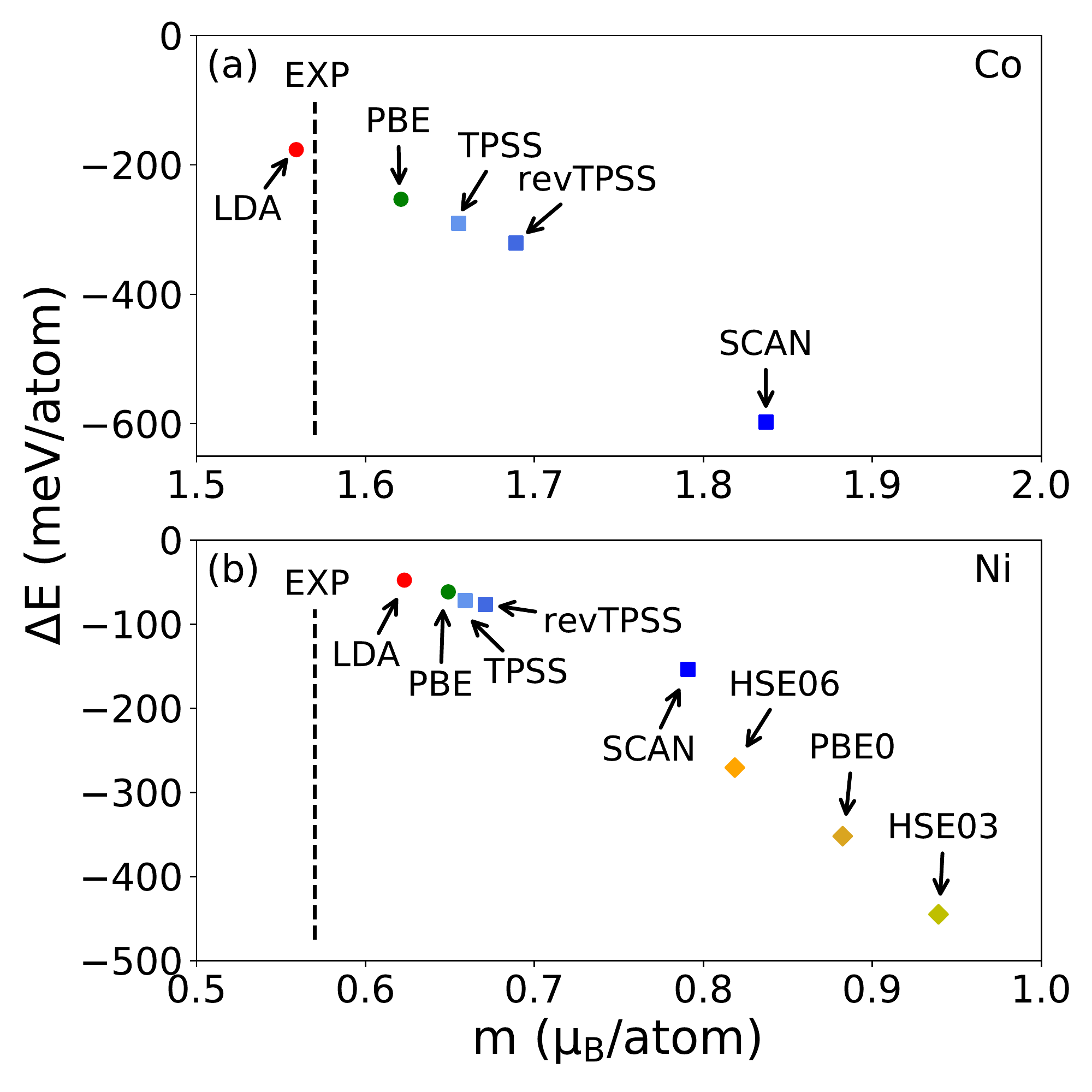}
\caption{Magnetic energy and spin magnetization per atom of hcp Co (top) and
fcc Ni (bottom) as obtained with different functionals. The LDA, PBE GGA and
meta-GGA calculations were done with VASP, and the hybrid functional
calculations were done with WIEN2k.}
\label{CoNi-energy}
\end{figure}

\begin{figure}
\includegraphics[width=0.9\columnwidth]{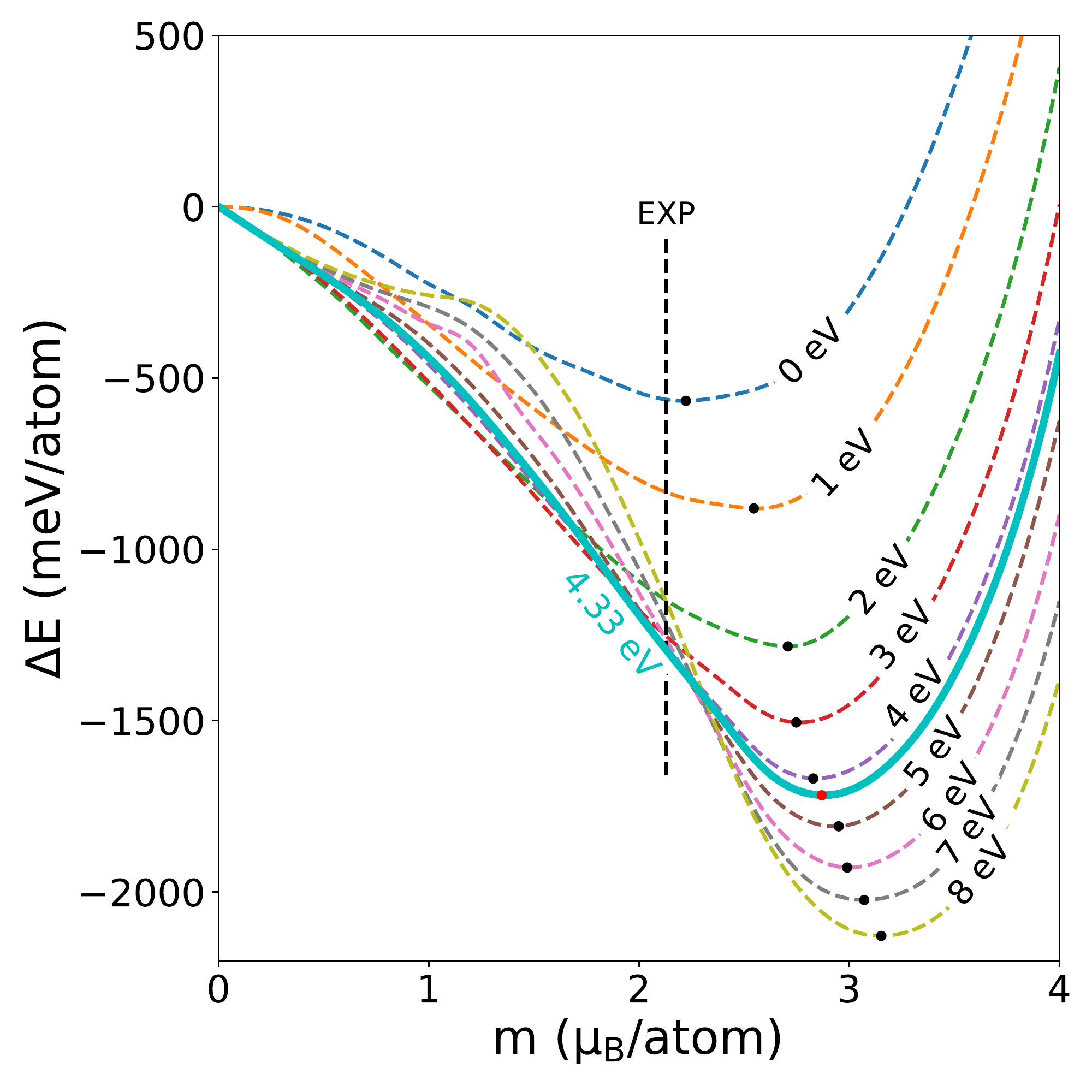}
\caption{Fixed spin moment energy as a function of spin magnetization
per atom for bcc Fe, as obtained with the PBE+$U$ method implemented
in WIEN2k.}
\label{dftU-Fe}
\end{figure}

\begin{figure}
\includegraphics[width=0.9\columnwidth]{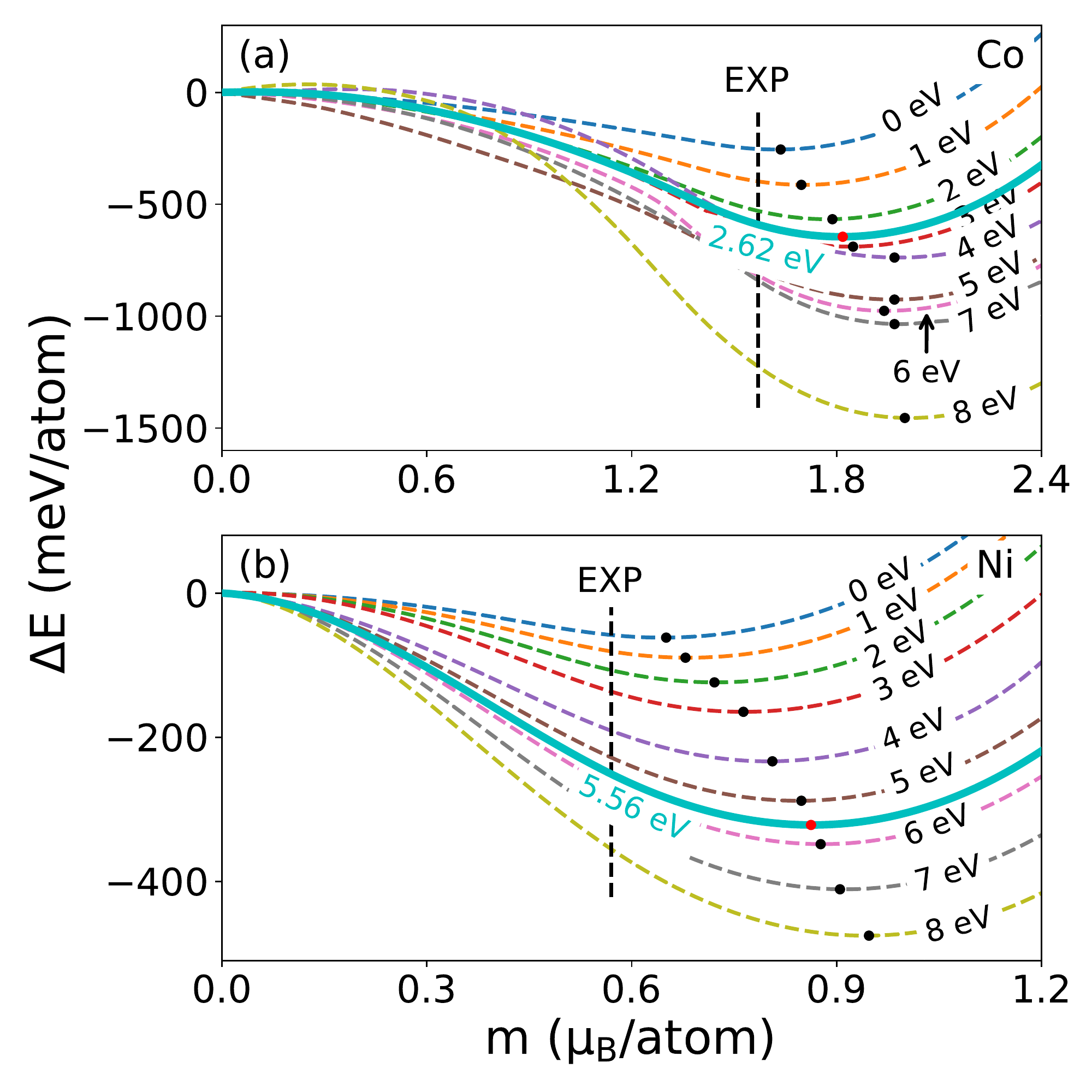}
\caption{Fixed spin moment energy as a function of spin magnetization
per atom for Co and Ni, as obtained with the PBE+$U$ method implemented
in WIEN2k.}
\label{dftU-CoNi}
\end{figure}

The experimental spin magnetizations of Fe, Co and Ni are
2.13 $\mu_B$, 1.57 $\mu_B$ and 0.57 $\mu_B$ per atom, respectively.
As noted previously, the LDA and PBE GGA functionals
provide values in good accord with experiment, while the
SCAN meta-GGA was found to provide values significantly higher than
experimental values.
\cite{ekholm,fu}
This is why SCAN is unable to describe the phase balance that
is central to the materials science of steel. \cite{fu}
Furthermore, although the overestimates of the magnetization with SCAN
amount to $\sim$25\% for Fe, the magnetic energies were found to be
greatly enhanced, by factors of two or more in these metals.
Moreover, as seen in
Fig. \ref{Fe-bcc-energy}, the SCAN meta-GGA behaves quite differently
than the other meta-GGA functionals, TPSS and revTPSS. Those functionals
also enhance the magnetic tendency of Fe relative to the PBE GGA, but to
a much smaller extent than SCAN. This is consistent with the previously
noted improved description of correlated materials using SCAN, and underscores
the difficulty of correctly describing both types of behavior with a
single method.

It is also important to note that the already overestimated
magnetic moments and energies with SCAN are further enhanced
with hybrid functionals,
which also describe the localized electron behavior
of correlated oxides, including FeO. \cite{perry,tran-06,alfredsson,rowan}
This enhancement is both in terms
of moments, as was previously noted,
\cite{paier,jang,janthon}
and also even more strongly in terms of magnetic energies.

We find similar results for Co and Ni. Spin magnetizations and
magnetic energies are shown in Fig. \ref{CoNi-energy}.
The main difference is that there is a greater relative
overestimation of the moments with the TPSS and revTPSS
meta-GGA functionals as compared to the PBE GGA.

As mentioned, another highly effective approach for localized systems
is the addition of a Hubbard $U$ correction, as in the widely
used LDA+$U$ and PBE+$U$ schemes.
The correction is designed to better distinguish orbitals, favoring
integer occupation, and thus correct delocalization errors in
standard LDA and GGA calculations.
In this sense these Hubbard corrections have physics related to that
of hybrid functionals, though with lower cost.
As shown in Figs. \ref{dftU-Fe} and \ref{dftU-CoNi},
they do not improve results for magnetism in Fe, Co and Ni.

\begin{figure}
\includegraphics[width=0.95\columnwidth]{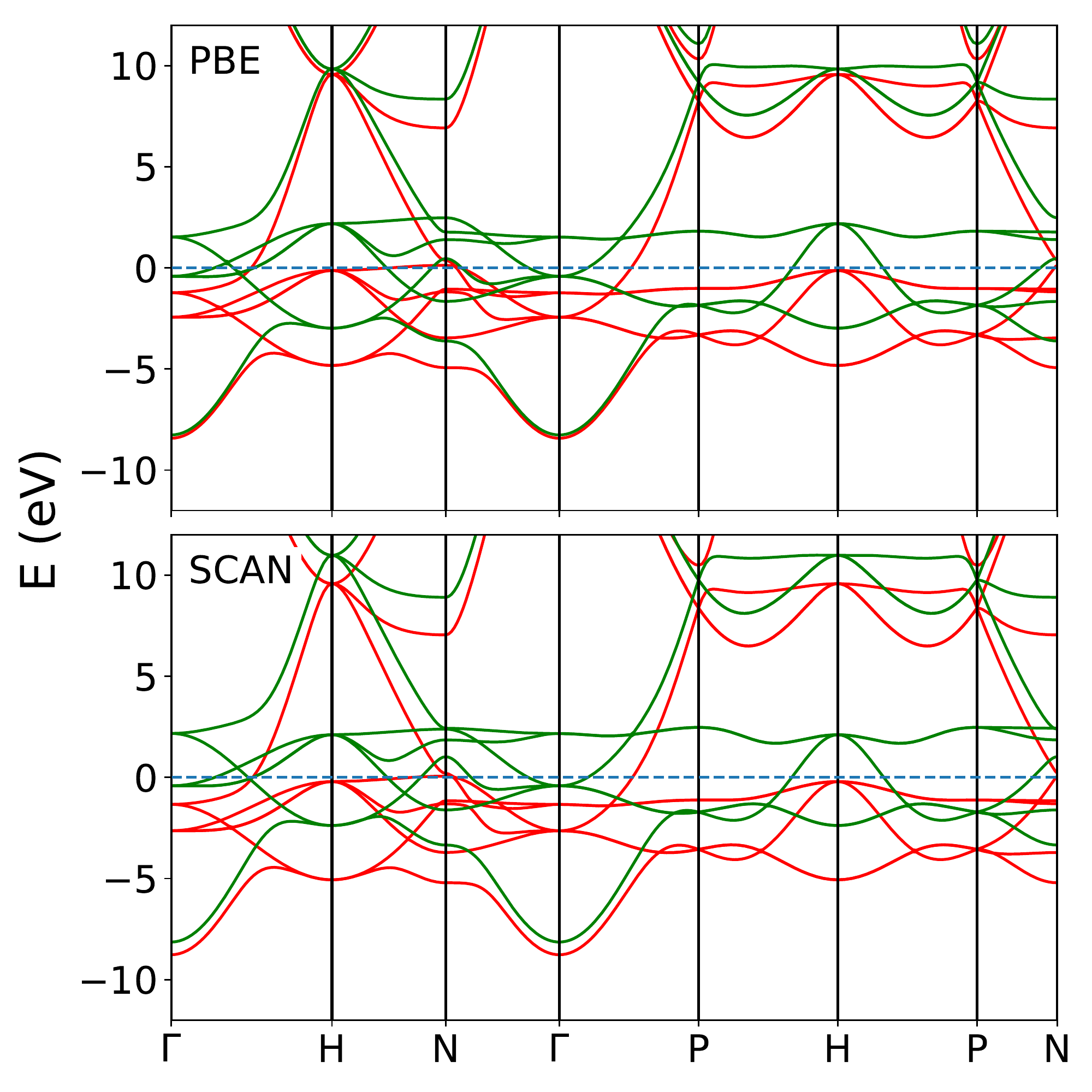}
\caption{Band structure of bcc Fe comparing PBE and SCAN,
as obtained from self-consistent calculations with VASP.
The Fermi level is at 0 eV and
majority (minority) spin are shown as light (dark) color.}
\label{band-Fe}
\end{figure}

It is important to note that PBE+$U$ strongly degrades agreement
with experiment for the values of $U$
obtained from linear response and
even for small values of $U$, e.g. 1 eV or 2 eV.
Thus the addition of a static Hubbard correction degrades
agreement with experiment for these metals even if the correction is made
small.

\begin{figure}
\includegraphics[width=0.95\columnwidth]{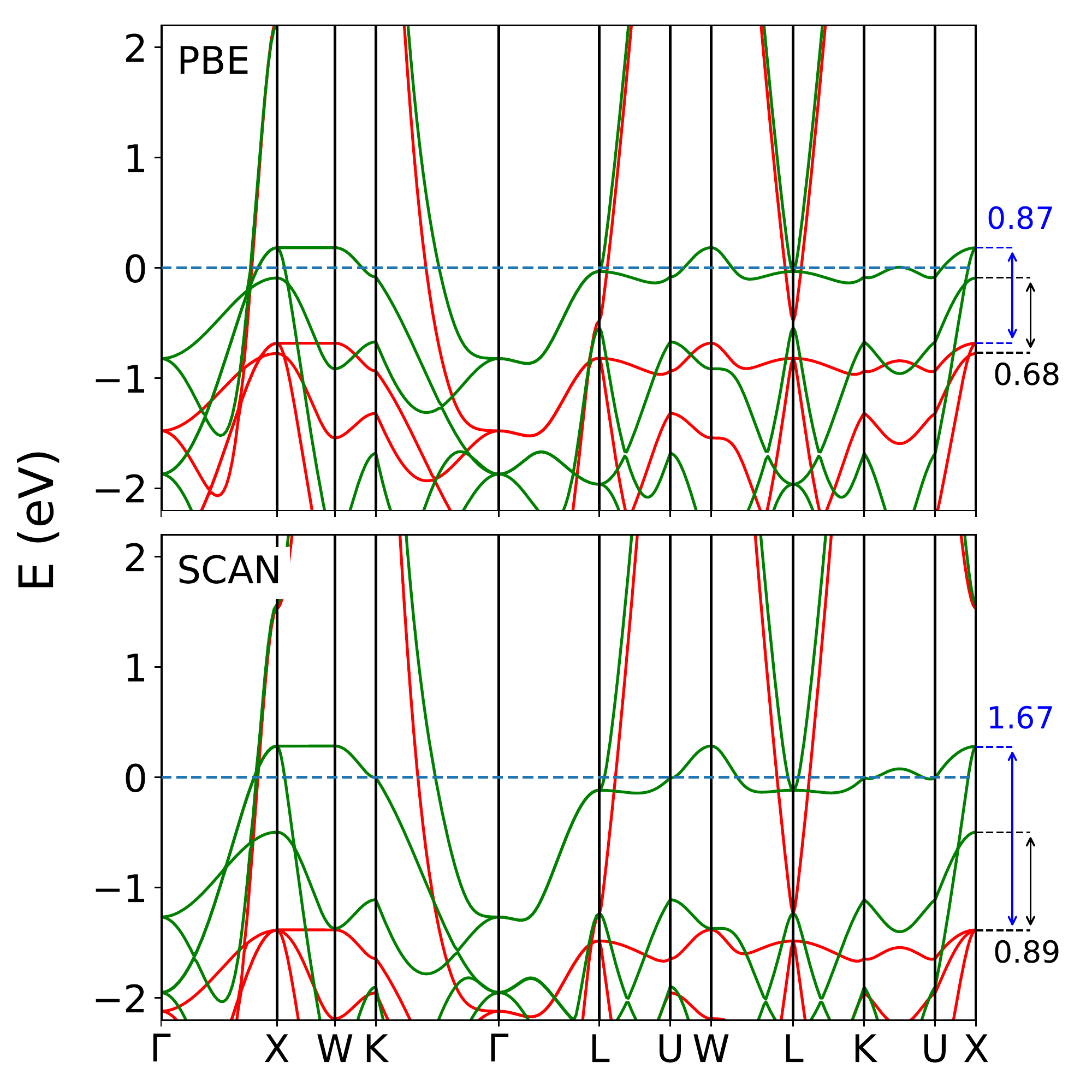}
\caption{Band structure of fcc Ni around the
Fermi level comparing PBE and SCAN,
as obtained from self-consistent calculations with VASP.
The Fermi level is at 0 eV and
majority (minority) spin are shown as light (dark) color.
The arrows and numerical values (in eV) give selected exchange splittings
at the $X$ point.}
\label{band-Ni}
\end{figure}

The Kohn-Sham eigenvalues in DFT do not correspond directly
with experimental excitation energies, and therefore care should be
taken in their interpretation. Nonetheless, within Kohn-Sham theory,
they do control the occupancy of the Kohn-Sham orbitals and therefore
the ground state properties, such as energy and magnetization.
For example, enhancing exchange splitting while keeping the band width
fixed will increase the magnetization.
As such, analysis of the band structures can provide useful insights
into the behavior of a functional.

Ekholm and co-workers observed that the $d$ states of
Fe are shifted to lower energy compared to the PBE GGA, due to
an increased exchange splitting, which disagrees with experiment.
\cite{ekholm}
This is closely connected with the larger magnetization, since if the
bands are relatively undistorted, the exchange splitting and
magnetization will be closely related.
Figs. \ref{band-Fe} and \ref{band-Ni}
show our band structures with PBE and SCAN for Fe and Ni.
As noted by Ekholm and co-workers,
the the band widths using SCAN and PBE are similar,
and the exchange splitting with SCAN is larger corresponding to larger
magnetizations.

\begin{figure}
\includegraphics[width=\columnwidth]{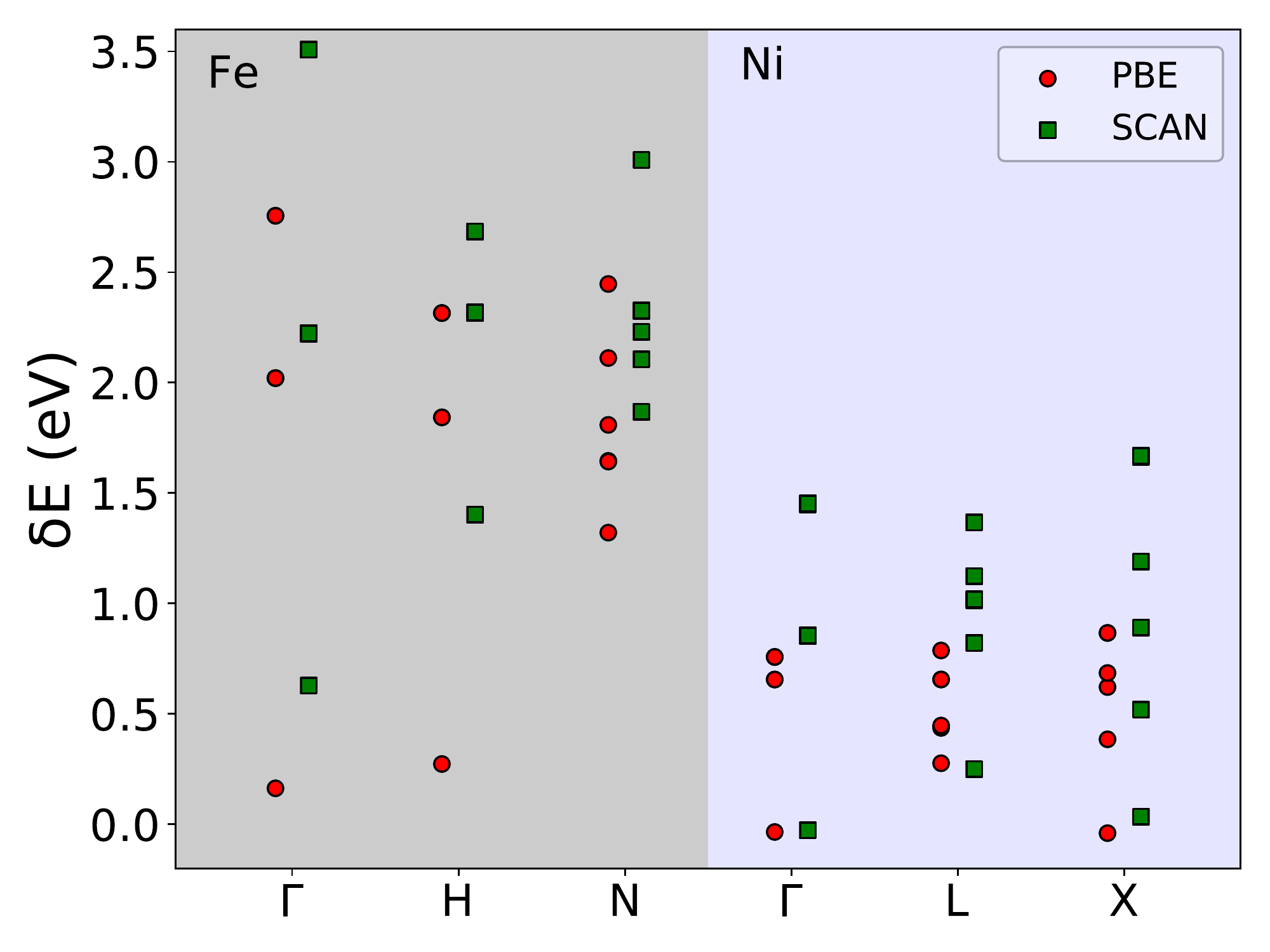}
\caption{Exchange splittings of the lowest six bands of bcc Fe
and fcc Ni from the band structures of Figs. \ref{band-Fe} and
\ref{band-Ni} at symmetry points.}
\label{exchange}
\end{figure}

In addition to the overall exchange splitting, there are
important differences between the PBE and SCAN band structures,
particularly for Ni.
The valence electronic structures of Fe and Ni can be roughly described as
consisting of bands originating in the 4$s$ orbital and the five 3$d$ states.
The lowest band at $\Gamma$ originates from the $s$ orbital and
disperses rapidly upwards mixing with the $d$ bands (note
that an $s$ band takes $p$-like
at the zone boundary, where it is above the Fermi level in these elements).
Thus the lowest six bands consist approximately of the $s$ band and the
$d$ band, which mix away from the $\Gamma$ point.
Fig. \ref{exchange} shows the exchange splittings of these bands for
Fe and Ni at symmetry points.

In Fe, with the LDA and PBE functionals, the
main $d$ band character is from $\sim$ -5 eV to +2 eV relative
to the Fermi energy $E_F$, the $d$ band occupancy is roughly seven electrons,
there is partial filling of all the $d$ orbitals, and
the band structure agrees well
with experiment aside from some renormalization of the
low energy bands, as is well
known. \cite{callaway,ackermann,santoni,schafer,walter}
It should be noted that an enhanced exchange splitting increases the
energy separation of occupied and unoccupied $d$ states, on average,
which is a way for favoring integer orbital occupations.

The electronic structure of ferromagnetic Ni has been extensively
studied.
It is roughly described as $d$ bands
and a partially filled $s$ band, with a $d$ band occupancy of roughly nine
electrons.
Differences between LDA and PBE calculations and
experiment are well established in Ni.
This includes a satellite feature observed in photoemission at $\sim$ 6
eV binding energy relative to the Fermi level,
\cite{davis}
which in any case is a many body effect that cannot be reproduced
by Kohn-Sham eigenvalues.
Additionally, Ni shows a considerably stronger renormalization of the $d$ bands,
as compared with Fe.
\cite{eastman,himpsel,liebsch,walter}
The $d$ band narrowing in Ni relative to LDA or PBE calculations is
accompanied by a decrease in the exchange splitting. These two
effects partially cancel as regards the spin magnetization,
so that while the exchange splitting is overestimated by a factor
of $\sim$2 in PBE calculations relative to experiment, the overestimate
of the spin magnetization is only $\sim$10\% (0.63 $\mu_B$ in PBE vs.
0.57 $\mu_B$ in experiment).

The electronic structure near the $X$ point, where an exchange split
$d$ band occurs near $E_F$, has been extensively studied.
\cite{eastman,himpsel,walter,tsui}
As seen in Fig. \ref{band-Ni}, the SCAN and PBE band structures are
remarkably different in this region. In particular, the PBE band
structure shows similar exchange splittings for the top $d$ bands.
The SCAN band structure on the other hand shows very different exchange
splittings for the top bands. The partially filled top $d$ band has an
exchange splitting of 1.67 eV at $X$, while the next lower band is exchange
split by 0.89 eV.
As seen in Fig. \ref{exchange}, the SCAN functional produces a much
larger range of exchange splittings in Ni than does the PBE functional.
The largest exchange splittings are in the topmost partially filled $d$
band. Thus SCAN much more strongly differentiates the orbitals in Ni
than PBE, again to the effect of favoring integer occupancy through
high exchange splittings of partially filled bands.

\section{Summary and Conclusions}

Calculations of the magnetic properties of Fe, Co and Ni with various
density functionals show that the SCAN functional is intermediate in
accuracy between the standard PBE functional, which provides a generally
good description of these materials, and approaches, such as PBE+$U$
and hybrid functionals that provide a poor description of these
ferromagnetic elements, but can describe more localized systems such as
Mott insulators.
SCAN is more different from the standard PBE than the earlier
TPSS and revTPSS meta-GGA functionals, and gives results
in worse agreement with experiment for the magnetism of Fe, Co and Ni.

The SCAN functional differentiates occupied from unoccupied states
more strongly than PBE, which is manifested in larger exchange
splittings in Fe and Ni, and also in the case of Ni in a greater difference
in the exchange splittings between different bands.
This can be understood as a greater tendency towards integer orbital
occupation, which is an important ingredient in describing atomic
systems and small molecules, and also is a key aspect of strongly
correlated materials such as Mott insulators.

Thus the challenge of developing a density functional approach that
can reliably and predictively treat transition metals and their compounds,
including both itinerant and localized systems remains to be solved.
This is an important problem not only from the point of view of physics,
but also in the context of materials science. This is because in materials
science the energies and resulting stabilities of different 
compounds and phases is important.
However, at present, as shown by the present results
functionals that describe the all the different phases that may be
important remain to be developed.
For example, none of the practical functionals tested
can give a good description of both metallic bcc and fcc Fe, and at the
same time Mott insulating oxides.
It may be extremely challenging to solve this problem because it
is unclear what semilocal information can be used to identify 
environments that lead to strong itinerancy from those that favor
localization.

\acknowledgments

We are grateful for helpful discussions with Jianwei Sun,
Fabian Tran and Samuel Trickey.
This work was supported by
the U.S. Department of Energy, Office of Science,
Basic Energy Sciences, Award Number DE-SC0019114.

\bibliography{scan}

\end{document}